\documentclass{andp2012}
\usepackage[english]{babel}

\usepackage{kantlipsum,cuted}

\usepackage[normalem]{ulem}
\usepackage{amsmath,amssymb}
\usepackage{multirow}
\usepackage{xcolor,soul}
\usepackage{srcltx}
\usepackage{hyperref,graphicx}

\title{Non-Hermitian Hartman effect}
\author[S: Longhi]{Stefano Longhi \inst{1,}\footnote{Corresponding author\quad E-mail:~\textsf{stefano.longhi@polimi.it}}}
\address[1]{Dipartimento di Fisica, Politecnico di Milano,Piazza L. da Vinci 32, I-20133 Milano, Italy and IFISC (UIB-CSIC), Instituto de Fisica Interdisciplinar y Sistemas Complejos, E-07122 Palma de Mallorca, Spain}
\shortauthors{S. Longhi }
\begin{abstract}
The Hartman effect refers to the rather paradoxical result that the time spent by a quantum mechanical
particle or a photon to tunnel through an opaque potential barrier becomes independent of barrier width for long barriers. Such an effect, which has been observed in different physical settings,  raised a lively debate and some controversies, owing to the correct definition and interpretation of tunneling times  and the apparent superluminal transmission. A rather open question is whether (and under which conditions) the Hartman effect persists for inelastic scattering, i.e. when the potential becomes non-Hermitian and the scattering matrix is not unitary. Here we consider tunneling through a heterojunction barrier in the tight-binding picture, where the barrier consists of a generally non-Hermitian finite-sized lattice attached to two semi-infinite nearest-neighbor Hermitian lattice leads. We derive a simple and general condition for the persistence of the  Hartman effect in non-Hermitian barriers, showing  that it can be found rather generally when non-Hermiticity arises from non-reciprocal couplings, i.e. when the  barrier displays the non-Hermitian skin effect, without any special symmetry in the system. 
\end{abstract}
\shortabstract
\begin{document}
\maketitle

\section{Introduction}
Tunnellng is one among the most  peculiar and popular phenomena predicted by quantum mechanics \cite{T1}. 
The question of how much time a quantum particle takes to tunnel
through a classically forbidden potential barrier is a rather old and somehow controversial topic in quantum physics \cite{T2,T3,T4,T5,T6,T7,T8,T9,T10,T11,T11b,T12,T13,T14,T15,T16,T16b}, owing to the difficulty to find a unique and convincing definition of the tunneling time \cite{T16,T17,T18,T19} and the apparent superluminal transmission that is found in some cases \cite{T9,T12,T13,T15,T16}. The Hartman effect \cite{T3},  i.e. the independence of the time delay for a quantum particle to tunnel across an opaque potential barrier, is perhaps one of the most intractable mysteries of tunneling that struggled physicists for more than four decades \cite{T7,T9,T10,T13,T16,T19b}. 
Measurements of tunneling times, including the demonstration of the Hartman effect and superluminal group delay times, have been reported in several experiments for electrons, photons and ultracold atoms\cite{T20,T21,T22,T23,T24,T25,T26,T27,T28,T28b,T29,T29b,T30,T31}. The relevance of tunneling times  has been discussed  in several areas of physics, including electron tunneling ionization in atoms \cite{T17,T28,T28b,T29,T29b}, photon tunneling  in microwave and optical barriers \cite{T20,T21,T22,T23,T24,T25,T26,T27}, tunneling of ultracold atoms \cite{T30,T31,T36} and electron tunneling in quantum superlattices and graphene \cite{T31b,T32,T33,T34,T35}.  
It is now well understood that the group delay time can result in apparent superluminal propagation or even negative group velocities, without violating causality and relativity \cite{T12,T16,T30,T37}. For recent works discussing the various controversies and interpretations of the Hartman effect see, for example, the review article \cite{T14} and the more recent works \cite{T16,T19b,T30,T37}. Most of previous results on the Hartman effect, with the exception of few works \cite{T38,T39,basta,T40,T41,T42,T43}, concern with non-dissipative (Hermitian) systems, where the dynamics conserves the norm of the wave function and the scattering matrix is unitary. However, in dissipative (open) quantum systems as well as in a variety of classical platforms such as photonic systems, mechanical metamaterials and topolectrical systems, scattering phenomena are described by effective non-Hermitian (NH) Hamiltonians, displaying non-unitary dynamics and possible singularities in the spectrum (see e.g. \cite{M1,M2,M3,M4,M5,M6} and references therein).  One class of such systems are those described by non-Hermitian Hamiltonians with parity-time (PT) symmetry, which were originally introduced in an attempt to  generalize the theory of quantum mechanics beyond the usual Hermiticity paradigm \cite{B1,B2}. PT symmetry  has become recently very popular and found several applications in photonics \cite{M2,M3,M4,M5}. In PT systems, non-Hermiticity basically arises from a complex potential with balanced dissipative (lossy) and amplifying (gain) regions. More recently, a great interest has been devoted toward a different type of non-Hermitian systems displaying the so-called non-Hermitian skin effect \cite{N1,N2,N3,N4,N5,N6,N7,N8,N9,N10,N11,N12,N13,N14,N15,N16,N17,N18,N19,N20,N21,N22,Referee1,Referee2,Referee3}, i.e. a strong sensitivity of the energy spectrum on the boundary conditions and the condensation of a macroscopic number of bulk states at the lattice edges under open boundary conditions \cite{N1,N2,N3,N4,N5,N6,N7,N8,N9}. Such non-Hermitian systems exhibit a rich and nontrivial band topology 
  \cite{N2,N10,N13,N15,N22}, which emerges rather generally from non-reciprocal hopping amplitudes induced by synthetic imaginary gauge fields \cite{N1,U1,U2,U3,U4,U5,U6}, rather than from complex on-site potentials. {\color{black} Lattices with effective non-reciprocal hopping amplitudes have been demonstrated in different physical systems, such as in photonic systems \cite{N18,Referee1,Referee2}, topolectrical circuits \cite{N20}, mechanical systems \cite{N19} and ultracold atoms \cite{Referee3}. In particular, the use of synthetic lattices in frequency domain \cite{Referee1}can realize rather arbitrary single-band NH Hamiltonians with tailored non-reciprocal hopping displaying arbitrary topological winding numbers}.\\ The impact of inelastic scattering on the tunneling times and Hartman effect has been investigated in few models \cite{T38,T39,T40,T41,T42,T43}, where non-Hermiticity was introduced via a complex potential term in the Hamiltonian. A general result is that, for a sufficiently strong absorption, the Hartman effect is washed out and the tunneling time turns out to depend on barrier width \cite{T38,T39,T40}; such a result is in agreement with some early  observations on photon tunneling in microwave experiments \cite{T39}. On the other hand, the Hartman effect persists in PT symmetric complex potential barriers \cite{T42,T43}, with alternating layers of absorption and gain,  suggesting that PT symmetry can play a main role in preserving the Hartman effect under non-Hermitian scattering processes \cite{T43}. However, it is a fully open question whether PT symmetry is a necessary requirement for the observation of the Hartman effect in non-Hermitian scattering processes, and whether the Hartman effect can be observed under different types of non-Hermiticity, involving imaginary gauge fields \cite{U1,U2,U3,U4,U5,U6}, so beyond the PT symmetry paradigm. \\
  In this work we address such main questions considering the problem of tunneling time for scattering in tight-binding lattices \cite{K1,K2,K3,K4,K5,K6,K7} where the barrier consists of a generally non-Hermitian finite-sized superlattice attached to two semi-infinite nearest-neighbor Hermitian lattice leads \cite{K3,K4,K7}. The main result is that under rather general conditions the Hartman effect persists in non-Hermitian models, also when the non-Hermitian barrier does not possess PT symmetry and there is the NH skin effect. The results are illustrated by considering tunneling across NH barriers described by the generalized Hatano-Nelson model and by a NH extension of the Rice-Mele model.

 \section{Transmission throughout a tight-binding non-Hermitian barrier: model and scattering analysis}
 We consider wave transmission through a non-Hermitian barrier on a one-dimensional (1D) tight-binding lattice, as schematically shown in Fig.1(a). Basically, we consider a tight-binding heterojunction where two semi-infinite tight-binding Hermitian chains (leads) with nearest neighbor hopping amplitude $\kappa$, providing the two scattering channels, are connected to the edge sites of a non-Hermitian barrier, which is described by a tight-binding superlattice containing $N$ unit cells and $M$ sites in each unit cell [Fig.1(a)]. Non-Hermiticity can be introduced by either (or both) complex on-site potentials $V_1$, $V_2$,..., $V_M$ in the sites $A_1$, $A_2$, .. $A_M$ of each cell or by non-reciprocal (asymmetric) hopping amplitudes. In the latter case the superlattice displays rather generally the NH skin effect \cite{N5}, although it can be found in some models with reciprocal hopping as well \cite{M4}. \par To study the scattering properties of the non-Hermitian barrier, let us consider a Bloch wave $\sim \exp(-iqn)$ with wave number $q$ and energy $E_0=2 \kappa \cos q$, with a positive group velocity $v_g=-(dE/dq)=2 \kappa \sin q$ (i.e. $0<q< \pi$), incident onto the barrier from the left side. Let us indicate by $H(\beta)$ the $(M \times M)$ matrix Hamiltonian in Bloch space of the superlattice in the barrier, with $\beta=\exp(ik)$. Clearly, in a lattice with short-range hopping the elements of the matrix Hamiltonian $H(\beta)$ are Laurent polynomials in $\beta$. The Bloch bands of the superlattice under periodic boundary conditions, 
 $E_{\alpha}(k)$ ($\alpha=1,2,..,M$), are the $M$ eigenvalues of the matrix Hamiltonian $H$ with $k$ real. Note that such energies are complex and, in the presence of the NH skin effect, they describe rather generally a set of closed loops $\mathcal{C}_1$, $\mathcal{C}_2$,..., $\mathcal{C}_M$ in complex energy plane [Fig.1(b)]. In analogy with the Hermitian case, we say that the barrier is opaque to an incident wave of energy $E_0$ whenever $E_0$ does not  belong to none of the loops $\mathcal{C}_1$, $\mathcal{C}_2$,..., $\mathcal{C}_M$. In this case, the wave function in the barrier region is described by a suitable superposition of evanescent waves, i.e. with complex Bloch wave numbers. However, it should be mentioned that, unlike in Hermitian barriers, owing to the NH skin effect the evanescent nature of the waves in the barrier does not necessarily imply that such waves are not propagative.\\
  To study wave scattering across the barrier,  let us consider the determinant
 \begin{equation}
 \Delta( \beta, E_0)=\det \left(E_0-H(\beta)  \right)
 \end{equation}
 which is a function of both $\beta$ and energy $E_0$ of the incoming wave. As a function of $\beta$, $\Delta(\beta,E_0)$ is a Laurent polynomial of the form 
 \begin{equation}
 \Delta(\beta,E_0)=\sum_{l=-s}^r R_l(E_0) \beta^l
 \end{equation}
  with some energy-dependent coefficients $R_l(E_0)$, where $s,r \geq 1$ are the largest orders of left ($s$) and right ($r$) hopping on the superlattice, respectively. 
  \begin{figure*}[htb]
\centerline{\includegraphics[width=17cm]{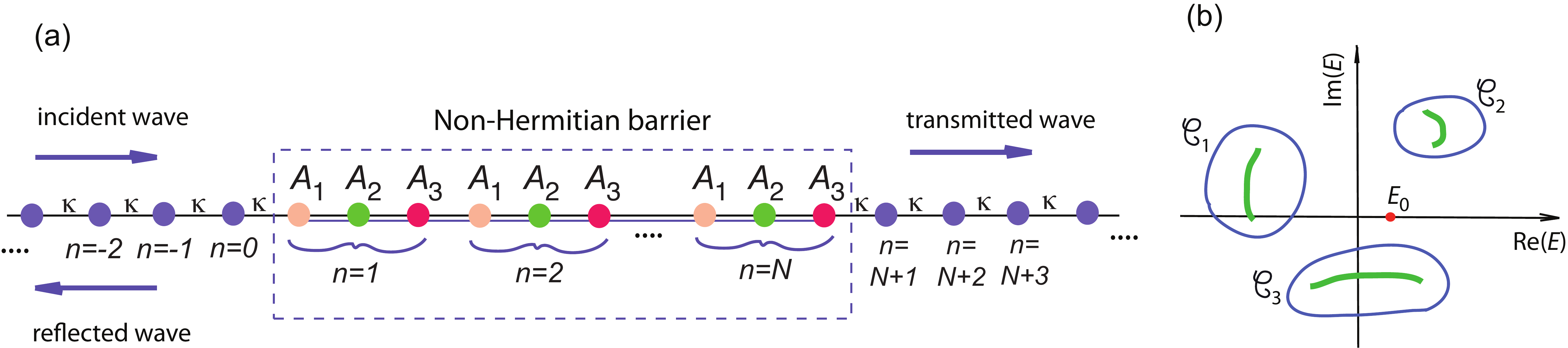}} \caption{ \small
(a) Schematic of wave scattering across a non-Hermitian barrier in a tight-binding heterojunction. The barrier consists of a chain of $N$ unit cells of a non-Hermitian superlattice, with $M$ sites $A_1$, $A_2$, ..., $A_M$ in each supercell ($M=3$ in the figure). The barrier is attached to two semi-infinite Hermitian tight-binding chains with nearest-neighbor hopping amplitude $\kappa$ (leads). The index $n$ denotes the number of the cell in the chain. (b) Energy spectrum of the superlattice under periodic boundary condition (solid curves). In the presence of non reciprocal hopping, the energy spectrum consists of a set of $M$ closed loops $\mathcal{C}_1$, $\mathcal{C}_2$, ..., $\mathcal{C}_M$ in complex energy plane ($M=3$ in the figure) and the system displays the NH skin effect. The superlattice barrier is opaque for the incoming wave of energy $E_0= 2 \kappa \cos q$ on the real energy axis whenever $E_0$ does not belong to none of the closed loops. The bold open arcs internal to the closed loops correspond to the energy spectrum of the NH superlattice under OBC. The condition $|\beta_{s}|<|\beta_{s+1}|$ for the observation of the Hartman effect physically implies that the energy $E_0$ of the incoming wave should not belong to the OBC energy spectrum.}
\end{figure*}  
The determinantal equation $\Delta(\beta,E_0)=0$ thus yields a set of $(s+r)$ roots $\beta_1$, $\beta_2$,  ..., $\beta_{r+s}$, which for an opaque barrier are  in modulus different than one. Therefore, in the barrier region the wave function is described by a superposition of evanescent waves which are either growing ($|\beta|>1$) or decaying $|\beta|<1$) as the cell number $n$ increases.
We order the roots such that
\begin{equation}
|\beta_1| \leq |\beta_2| \leq ... \leq |\beta_{s+r}|
\end{equation}
and indicate by $U^{(1)}$, $U^{(2)}$,..., $U^{(s+r)}$ the associated eigenvectors of $H$, i.e. $H(\beta_l)U^{(l)}=E_0 U^{(l)}$. For a barrier of width $N$ larger than $(r+s)$, the scattering solution $\psi_n$ to  the time-independent  Schr\"oedinger equation corresponding  to the incoming wave of energy $E_0$ from the left side of the barrier can be written in the form
\begin{equation}
\psi_n= \left\{
\begin{array}{lr}
\exp(-iqn)+ \mathcal{R}(q) \exp(iqn) & n \leq 0 \\
\sum_{l=1}^{s+r}G_l U^{(l)} \beta_l^n & 1 \leq n \leq N \\
\mathcal{T}(q) \exp[-iq(n-N-1)] & n \geq N+1\\
\end{array}
\right.
\end{equation}
 where $\mathcal{R}(q)$, $\mathcal{T}(q)$ are the spectral reflection and transmission coefficients (for left incidence side) and $G_l$ the amplitudes of evanescent waves in the barrier region. Note that $\psi_n$ is a scalar for $n<1$ and $n>N$, while it is a vector of size $M$ in the barrier superlattice, containing the wave amplitudes in each site $A_1$, $A_2$, ..., $A_M$ of the $n$-th cell ($n=1,2,...,N$) of the barrier superlattice. The values of the $(s+r+2)$ variables $\mathcal{R}(q)$, $\mathcal{T}(q)$, $G_1$, $G_2$,..., $G_{r+s}$ are obtained by imposing the matching conditions at the two junctions. These include the two matching equations at sites $n=0$ and $n=N+1$, in the leads, given by
 \begin{eqnarray}
 E_0 (1+\mathcal{R})=\kappa [ \exp(iq)+\mathcal{R} \exp(-iq)] +\kappa  \sum_{l=1}^{r+s} G_l U^{(l)}_1 \beta_l \\
 E_0 \mathcal{T} = \kappa \mathcal{T} \exp(-iq)+ \kappa \sum_{l=1}^{r+s} G_l U^{(l)}_M \beta_l^N,
 \end{eqnarray}
 and other $(r+s)$ equations for the first $s$ sites in the superlattice at the left edge, and the last $r$ sites in the superlattice at the right edge. Such equations are linear in the amplitudes $G_l$ with coefficients that contain the powers $\beta_{l}, \beta_{l}^2,..., \beta_{l}^{r+s}$ and $\beta_{l}^{N-s-r+1}, \beta_{l}^{N-s-r+2},..., \beta_{l}^N$. The explicit form of such equations is given in the Appendix A for the special case of $M=1$, while an illustrative example for $M=2$ is given in Sec.4.2. The resulting system is a set of $(s+r+2)$ linear and inhomogenous equations in the $(s+r+2)$ unknown variables $\mathcal{R}(q)$, $\mathcal{T}(q)$, $G_1$, $G_2$,..., $G_{r+s}$, which can be solved to determine the spectral transmission amplitude $\mathcal{T}(q)$.
 
 \section {Tunneling phase time and the non-Hermitian Hartman effect} 
 The Hartman effect refers to the independence of the tunneling phase time (also referred to as the group delay time or Wigner time) on the barrier width for long opaque potential
barriers \cite{T3,T9,T15,T16}. Here we do not enter into the rather longstanding and controversial question whether the phase time does provide an accurate estimate of the traversal time (see e.g. \cite{T15,T16} and reference therein), however we mention that 
 such an independency of barrier width can be found for other types of mean tunnelling times \cite{T15} and that, according to Ref.\cite{T16}, the group delay has a clear physical
significance despite its interpretation as a traversal time cannot be justified. {\color{black}{ When dealing with wave tunneling in a non-Hermitian barrier, the phase time keeps its clear physical meaning as the delay time of the peak of a spatially-broad wave packet to escape from the barrier. This time can be readily measured in a tunneling experiment based, for example, on light propagation across NH photonic barriers \cite{T25,T26}. Other tunnelling times, such as the dwell time, could not be readily extended to non-Hermitian barrier tunneling or could not be defined uniquely. For example, in \cite{basta}  it was shown that different dwell times can be introduced in non-Hermitian tunneling, and that one of them can become a complex number, with a nontrivial physical meaning. }}\\
 In this section we aim at calculating the tunneling phase time in the scattering process of Fig.1(a), i.e. across a NH potential barrier in the tight-binding picture.
 Let us consider  a spatially-broad wave packet incident onto the NH barrier from the left side with a spectral amplitude $F(\delta)$ very narrow at around the mean Bloch wave number $\delta=q$. For $n \leq 0$ the wave function can be written as the sum of the incident $ \psi_n^{(inc)}(t)$ and reflected $ \psi_n^{(ref)}(t)$ wave packets i.e.  $ \psi_n(t)=\psi_n^{(inc)}(t)+ \psi_n^{(ref)}(t)$ with
 \begin{eqnarray}
  \psi_n^{(inc)}(t) & = & \int d \delta F( \delta) \exp[-i \delta n-iE(\delta)t] \\
  \psi_n^{(ref)}(t) & = & \int d \delta \mathcal{R}(\delta) F( \delta) \exp[i \delta n-iE(\delta)t]
 \end{eqnarray}
 while for $n>N$ the transmitted wave packet, on the right hand side of the barrier, reads
  \begin{equation}
 \psi_n^{(trans)}(t)=\int d \delta F( \delta) \mathcal{T}(\delta) \exp[-i\delta (n-N-1)-iE(\delta)t].
 \end{equation}
After letting $\mathcal{T}(q)=|\mathcal{T}(q)| \exp[i \varphi_t(q)]$, using the method of stationary phase \cite{T9,T15,T16} one can calculate the phase (group delay) time, from $n=0$ to $n=N+1$, which reads
\begin{equation}
\tau=\frac{(d \varphi_t /dq)}{ (dE/dq)}=-\frac{(d \varphi_t /dq)}{ 2 \kappa \sin q}. \label{phasetime}
\end{equation}
Therefore, the tunneling phase time $\tau$ is basically established by the derivative of the phase of the spectral transmission amplitude.
To establish whether the phase time becomes independent of $N$ (the barrier width) in the large $N$ limit, we use a main result, proved in Appendix B, for which in the large $N$ limit the form of the spectral  transmission coefficient $\mathcal{T}(q)$ reads
\begin{equation}
\mathcal{T}(q)=\tilde{\mathcal{T}}(q) \beta_s^N \label{theorem}
\end{equation}
where $\tilde{\mathcal{T}}(q)$ is independent of $N$. The above result holds provided that the inequality $|\beta_s|< | \beta_{s+1}|$ strictly holds, where $\beta_1$, $ \beta_2$, ..., $\beta_{r+s}$ are the $(s+r)$ roots of the determinantal equation $\Delta(\beta,E_0)=0$ ordered according to Eq.(3). Interestingly, the condition $|\beta_s|< | \beta_{s+1}|$ for the validity of Eq.(11) implies that the energy $E_0$ of the incoming wave does not belong to the energy spectrum of the NH superlattice under open boundary conditions (OBC) (see \cite{
  N11, N15, N17,healing} and Appendix A).

From Eqs.(\ref{phasetime}) and (\ref{theorem}) it readily follows that the phase time reads
\begin{equation}
\tau=-\frac{(d \tilde{\varphi} /dq)}{ 2 \kappa \sin q}-N \frac{(d {\varphi_{\beta_s}} /dq)}{ 2 \kappa \sin q}  
\end{equation}
where $\tilde{\varphi}(q)$ and $\varphi_{\beta_s}(q)$ are the phases of $\tilde{\mathcal{T}}(q)$ and $\beta_s(q)$, respectively. Therefore, we conclude that the necessary and sufficient condition for the Hartman effect to occur in the tunneling across the NH barrier is that the phase of the root $\beta_s$ does not depend on $q$ in the neighbor of the carrier Bloch wave number $q$ of the incoming wave packet, i.e. that the stationarity condition $(d \varphi_{\beta_s} /dq)=0$ is satisfied. This occurs, for instance, whenever $\beta_s(q)$ is a real number, and this condition can occur under rather broad selection of parameters, even without any symmetry in the system. For example, let us consider the case $M=1$, so that $H(\beta)$ is a scalar and given by the Laurent polynomial
\begin{equation}
H(\beta)=\frac{t_{-s}}{ \beta^{s}}+\frac{t_{-s+1}}{ \beta^{s-1}}+...+ t_0 + t_1 \beta+ ...+ t_r \beta^r
\end{equation}
where $t_{\mp l}$ are the left or right hopping amplitudes among sites
distant $|l|$ in the lattice and $s, r \geq 1$ are the largest orders of
left or right hopping. {\color{black}{Such a single-band NH Hamiltonian could be implemented, for example, in photonics using a synthetic lattice in the frequency domain, where a ring resonator undergoes simultaneous phase and amplitude modulations that control the left and right hopping amplitudes in the lattice \cite{Referee1}}.}
The system displays the NH skin effect whenever $|t_{-l}| \neq |t_l |$ for some $l$. The $\beta_l$ values are the roots of the following polynomial
\begin{equation}
t_r \beta^{s+r}+t_{r-1} \beta^{s+r-1}+...+(t_0-E_0) \beta^s+...+t_{-s+1} \beta+t_{-s}=0.
\end{equation}
If the hopping amplitudes $t_l$ are real, a sufficient condition for the roots of the polynomial to be real and distinct (so that $\beta_s$ is real and $|\beta_s|$ is strictly smaller than $|\beta_{s+1}|$) is given by a general theorem of linear algebra \cite{algebra}, requiring some constraints on the hopping amplitudes. Some illustrative examples of NH
 barrier models displaying real $\beta_s$ values, without requiring any special symmetry (such as PT symmetry) in the system, are presented in the next section.\\
A final comment is in order concerning the observability of the NH Hartman effect in an ideal wave packet tunneling experiment. Since we are dealing with a NH barrier, besides scattering states with real energy $E_0= 2 \kappa \cos q$ the heterojunction system of Fig.1(a) could sustain a set of bound (normalizable) states, localized near the barrier region, with complex eigenenergies $E_1$, $E_2$,... The energies of such bound states are given by $E_1= 2 \kappa \cos q_1, E_2= 2 \kappa \cos q_2$, ..., where $q_1$, $q_2$, ... are the poles of the spectral transmission and reflection amplitudes $\mathcal{T}(q)$, $\mathcal{R}(q)$ in the complex $q$ plane with ${\rm Im} (q)<0$. The bound states are unstable if the imaginary part of their energy is positive. In this case, i.e. in presence of unstable states, the Hartman effect could be hard to be observed (or could be observed only transiently), since any noise in the system can secularly amplify the unstable bound states, preventing the observation of the tunneled wave packet. On this issue, see for instance Ref. \cite{unstable}. We note that, if the energy spectrum of the potential barrier under OBC (i.e. when not attached to the two left and right leads) is real or all energies are stable (i.e. with negative imaginary parts, like in purely dissipative systems), then it is likely that unstable bound states will not arise when we attach the barrier to the left and right Hermitian leads, a result which is exact in the  $\kappa \rightarrow 0$ limit. In the following, we will not consider NH barrier potentials that display unstable bound states, or assume that possible unstable bound states  have a small imaginary part of the eigenenergy so as not to prevent the observation of wave packet tunneling over a limited temporal window.

 \begin{figure}[htb]
\centerline{\includegraphics[width=8.6cm]{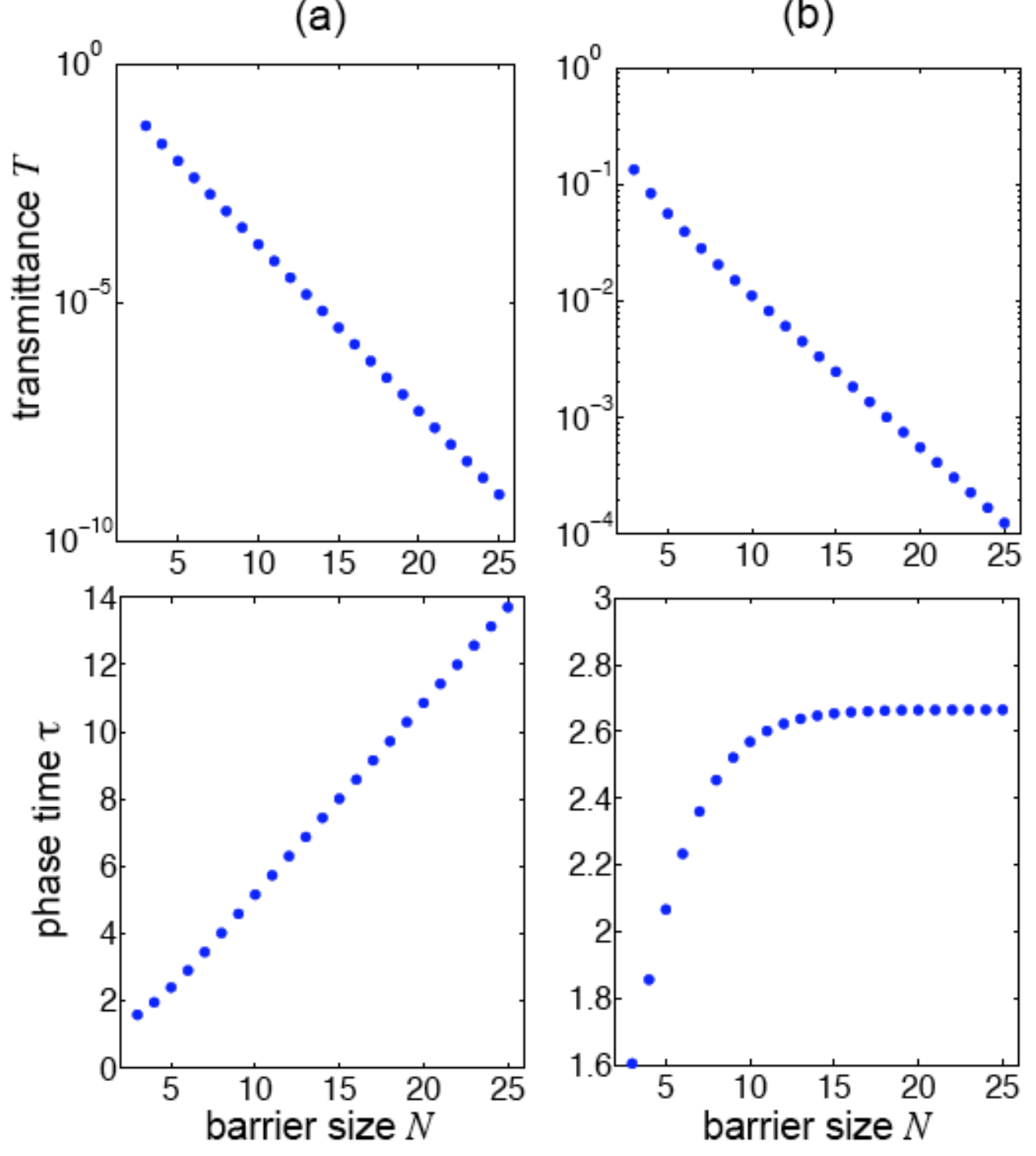}} \caption{ \small
Behavior of the transmittance $T=|\mathcal{T}|^2$ on a log scale (upper panels), and of tunneling phase time $\tau$ (lower panels) versus the barrier size $N$ in two different types of NH barriers with nearest-neighbor hopping. (a) A dissipative barrier with complex on-site potential ($\kappa=t_{-1}=t_1=1$, $t_0=2.1-0.2i$); (b) The NH Hatano-Nelson barrier with non-reciprocal hopping and real on-site potential ($\kappa=t_{-1}=1$, $t_0=1.85$, $t_1=0.8$). The incident wave has a Bloch wave number $q=\pi/2$, corresponding to the energy $E_0=2 \kappa \cos q=0$. Note that the Hartman effect, i.e. the independence of the phase time $\tau$ on barrier size in the large $N$ limit, is found in the second model solely.}
\end{figure} 

 \begin{figure}[htb]
\centerline{\includegraphics[width=8.6cm]{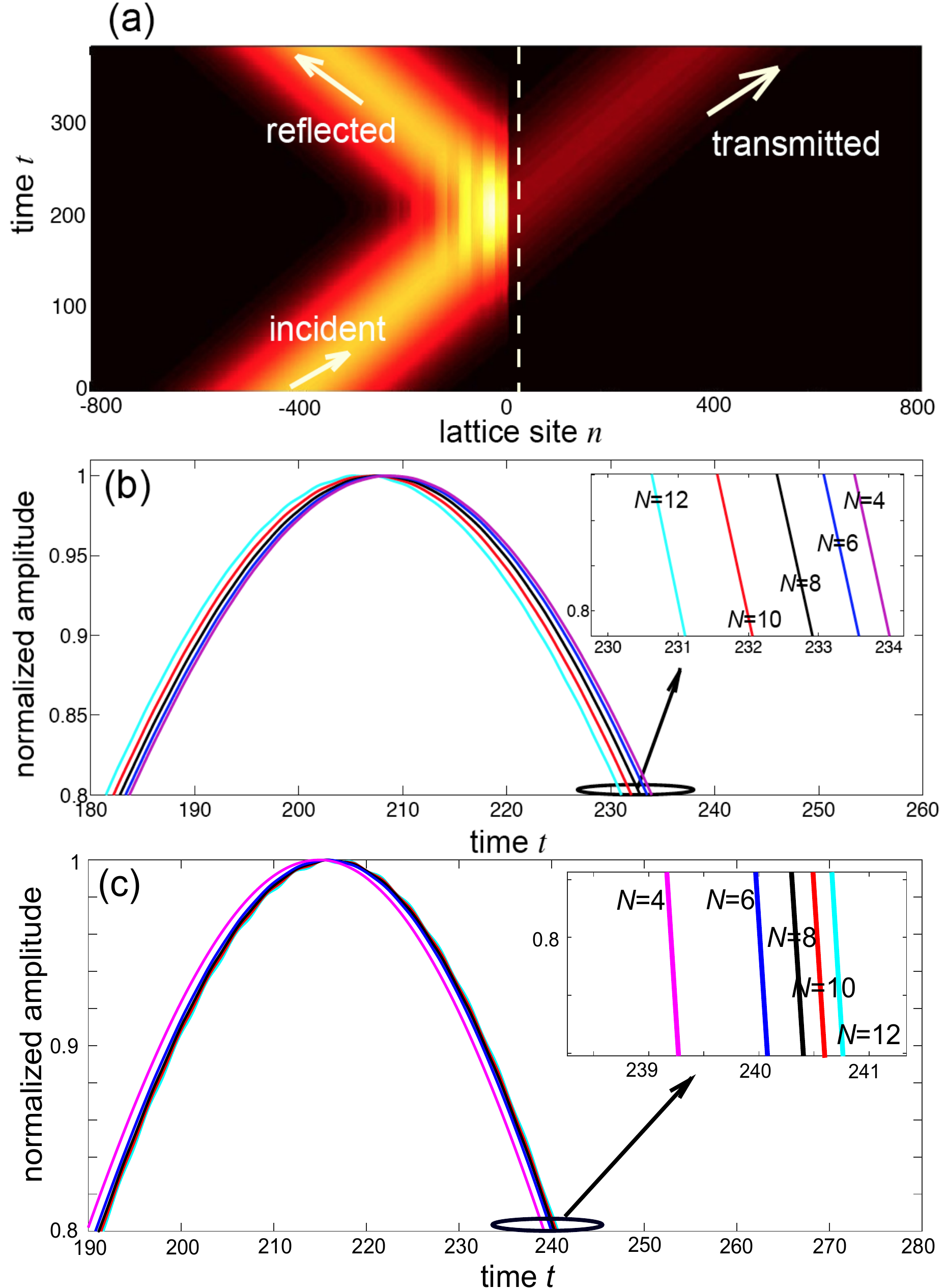}} \caption{ \small
Scattering dynamics at a NH barrier with non-reciprocal nearest neighbor hopping amplitudes [$\kappa=t_{-1}=1$, $t_{0}=1.85$, $t_1=0.8$ as in Fig.2(b)]. (a) Space-time scattering dynamics of an incident Gaussian wave packet for barrier width $N=6$. The input wave function is $\psi_n(0)=\exp[-(n+400)^2/w_0^2-iqn] $ with carrier Bloch wave number $q=\pi /2$ and size $w_0=150$. (b) Temporal behavior of the wave function amplitude $| \psi_n(t)|$ in the site $n=20$, after the barrier [indicated by the dashed vertical curve in (a)], normalized to its maximum value, for a few increasing values of the barrier size $N$ ($N=4,6,8,10,12$). {\color{black}{The barrier is located between $n=1$ and $n=N$.}}  As shown in the inset, as $N$ increases the rate $\Delta \tau / \Delta N$ of the waveform time advancement $\Delta \tau$ reaches a value close to 0.5, indicating that the tunneling phase time becomes nearly independent of the barrier size $N$ for large $N$. {\color{black}{(c) Same as (b), but for parameter values  $\kappa=t_{-1}=t_1=1$, $t_0=2.1-0.2i$ [as in Fig.2(a)]. In this case there is not any advancement of the waveform as the barrier $N$ is increased.}}}
\end{figure} 

\section{Illustrative examples and numerical simulations}
We illustrate the general results on the NH Hartman effect, presented in the previous section, by considering two prototypal models of NH tight-binding barriers, corresponding to $M=1$ (the generalized Hatano-Nelson model) and $M=2$ (the NH Rice-Mele model).
\subsection{The generalized Hatano-Nelson model}
Let us assume that the unit cell of the barrier comprises a single site, i.e. $M=1$, and that we have only nearest-neighbor hopping $t_{-1}, t_{1}$ in the lattice ($r=s=1$), so that $H(\beta)=t_{-1}/ \beta+ t_0 +t_1 \beta$. This corresponds to the clean Hatano-Nelson model \cite{N2,U1}, which does not possess any symmetry. The roots $\beta_1$, $\beta_2$ to the determinantal equation $H(\beta)-E_0=0$ read
\begin{equation}
\beta_{1,2}= \frac{E_0-t_0}{2t_1} \pm \frac{1}{t_1} \sqrt{\left( \frac{E_0-t_0}{2} \right)^2 -t_1t_{-1}}
\end{equation}
with $E_0=2 \kappa \cos q$. Then the condition that the phase of $\beta_s=\beta_1$ does not depend on $q$ can be met provided that the $\beta ^{\prime}$s roots are real. This excludes the case where the on-site potential $t_0$ is complex, such as in a barrier with dissipation. However, when the on-site potential energy $t_0$ and $t_1 t_{-1}$ are real numbers, the condition for the observation of the NH Hartman effect is satisfied for $(E_0-t_0)^2/4-t_1t_{-1}>0$.  In particular, for an incident wave with zero energy $E_0=0$ at the band center, corresponding to a Bloch wave number $q=\pi/2$, 
the NH Hartman effect is observed provided that $t_0^2>4 t_1 t_{-1}$. {\color{black}{In this case, in order to avoid the appearance of unstable modes we assume $t_1t_{-1}>0$, corresponding to an entirely real energy spectrum of the Hatano-Nelson Hamiltonian under open boundary conditions.}}
 As an example, Fig.2 shows the spectral transmittance $T=|\mathcal{T}|^2$ and tunneling phase time $\tau$ versus barrier size $N$, as obtained by numerically solving the matching conditions at the junctions [Eqs.(A.11-A.19) in Appendix A], for an incidence wave with zero energy ($q= \pi/2, E_0=0$) in two types of NH barrier models. In the former model [Fig.2(a)]  we have a dissipative on-site potential barrier with symmetric hopping amplitudes, which does not display the NH Hartman effect. In the latter case we have a NH barrier with asymmetric hopping amplitudes and real on-site potential [Fig.2(b)], which clearly shows the NH Hartman effect, indicated by the saturation of the tunneling phase time $\tau$ as $N$ is increased.\\
We checked the occurrence of the Hartman effect in the Hatano-Nelson barrier model with non-reciprocal hopping and real on-site potential by direct numerical simulations of the time-dependent Schr\"odinger equation on the lattice. A spatially-broad Gaussian wave packet with carrier Bloch wave number $q=\pi/2$ is injected to the barrier from the left side, and the peak delay of the transmitted wave packet is  detected after the scattering event. Figure 3(a) depicts on a pseudocolor map a typical temporal evolution of the normalized occupation amplitudes $| \psi_n(t) | / \sqrt{\sum_n | \psi_n(t)|^2}$ for a barrier size $N=6$ and for the same parameter values as in Fig.2(b).  The initial wave packet, at time $t=0$, is well localized on the left side of the barrier with a broad Gaussian distribution of size $w_0=150$ (much larger than the barrier width to limit wave packet distortion effects \cite{T16,T27}) and carrier Bloch wave number $q= \pi/2$, corresponding to the energy $E_0=0$. Figure 3(b) shows the temporal behavior of the wave function amplitude $| \psi_n(t)|$ in the site $n=20$ after the barrier [indicated by the dashed vertical curve in Fig.3(a)], normalized to its maximum value, for a few increasing values of the barrier size $N$ ($N=4,6,8,10,12$).  {\color{black}{The delay time $\tau_d(N)$ of the transmitted wave packet at the reference site $n=20$, for a barrier of size $N$, can be written as $\tau_d(N)=\tau_{d}^{(1)}+\tau_d^{(2)}$, where $\tau_{d}^{(1)}$ is the delay time due to the crossing of the barrier and $\tau_{d}^{(2)}$ is the traversal time in the leads, where the wave packet propagates at the group velocity $v_g=2 \kappa \cos q=2$. In the presence of the Hartman effect, i.e. independence of $\tau_{d}^{(1)}$ on barrier size $N$ for large $N$, as we increase the barrier size $N$ by $\Delta N$, $\tau_{d}^{(1)}$ does not change while $\tau_{d}^{(2)}$ clearly decreases by $\Delta N / v_g= \Delta N /2$, since the spatial distance that the wave packet spends in the leads is reduced by $\Delta N$. This means that, indicating by $\Delta \tau=\tau_d(N)-\tau_d(N+\Delta N)$ the time advancement of the peak of the wave packet, at the reference site $n=20$, as we increase the barrier width from $N$ to $N+\Delta N$, in the presence of the Hartman effect the ratio $\Delta \tau / \Delta N$ should reach the value $\Delta \tau / \Delta N \simeq 1/2$, independent of $N$}}. In Fig.3(c) we can clearly see a slight advancement $\Delta \tau$ in time  of the peak of the waveform as $N$ is increased, with a rate $\Delta \tau / \Delta N$  that reaches an asymptotic value close to $0.5$ as $N$ is increased [see the inset in Fig.3(b)], which is thus a signature of the independence of tunneling phase time on barrier width. {\color{black} For parameter values where the phase of $\beta_s(q)$ does not show any stationary point, for example in the dissipative barrier of Fig.2(a), there is not any advancement of the waveform as $N$ is increased [see Fig.3(c)], indicating the absence of the Hartman effect. }\\

\begin{figure}[h]
\centerline{\includegraphics[width=8.6cm]{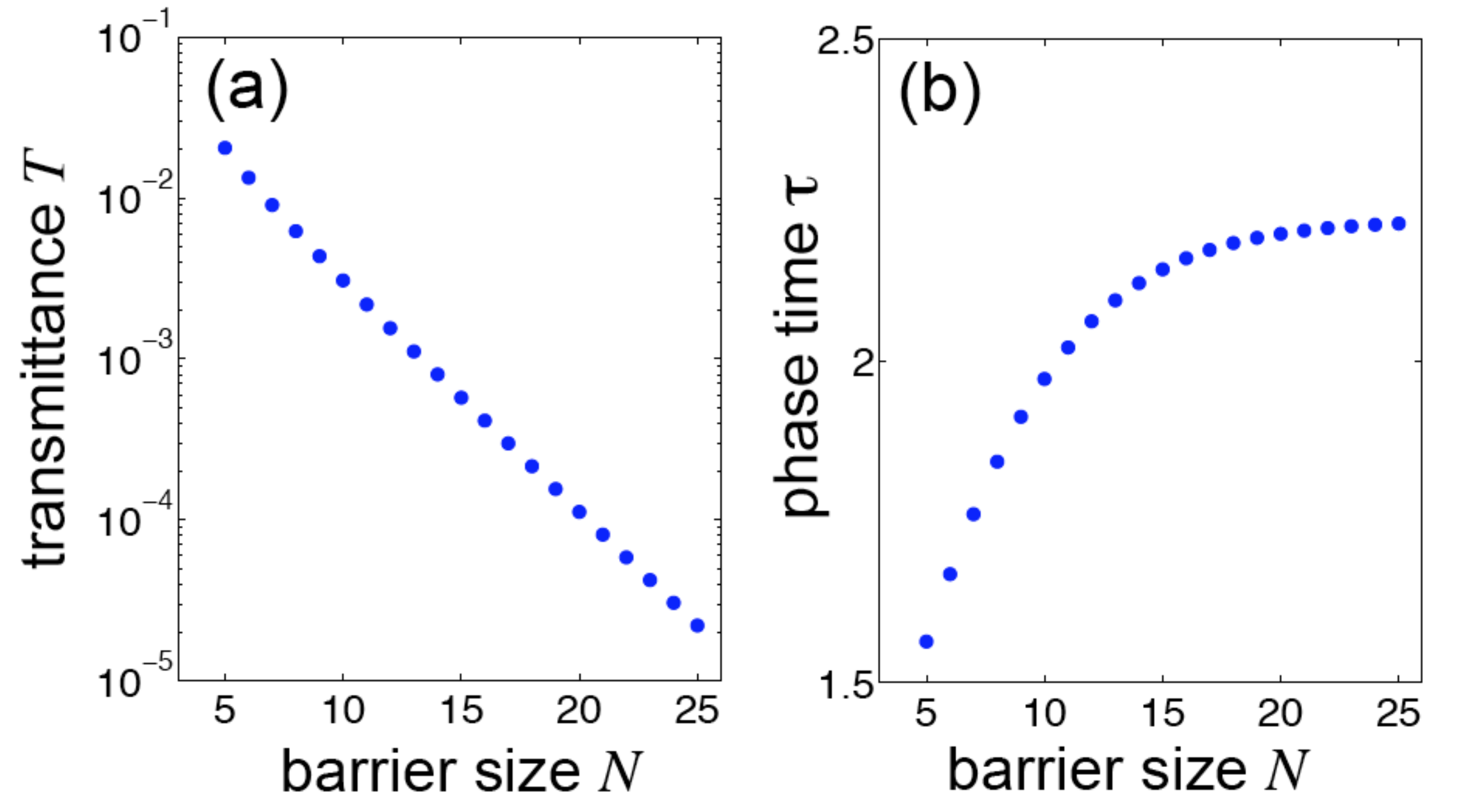}} \caption{ \small
Behavior of the transmittance $T=|\mathcal{T}|^2$ on a log scale (a) and tunneling phase time $\tau$ (b) versus the barrier size $N$ in the NH generalized Hatano-Nelson barrier with nearest and next-to-nearest hopping amplitudes 
$\kappa=1$, $t_{-2}=0.275$, $t_{-1}=-0.8667$, $t_{0}=-2.3627$, $t_1=-0.9184$ and $t_2=0.25$. The incident wave has a Bloch wave number $q=\pi/2$, corresponding to the energy $E_0=2 \kappa \cos q=0$.}
\end{figure} 
 \begin{figure}[h]
\centerline{\includegraphics[width=8.6cm]{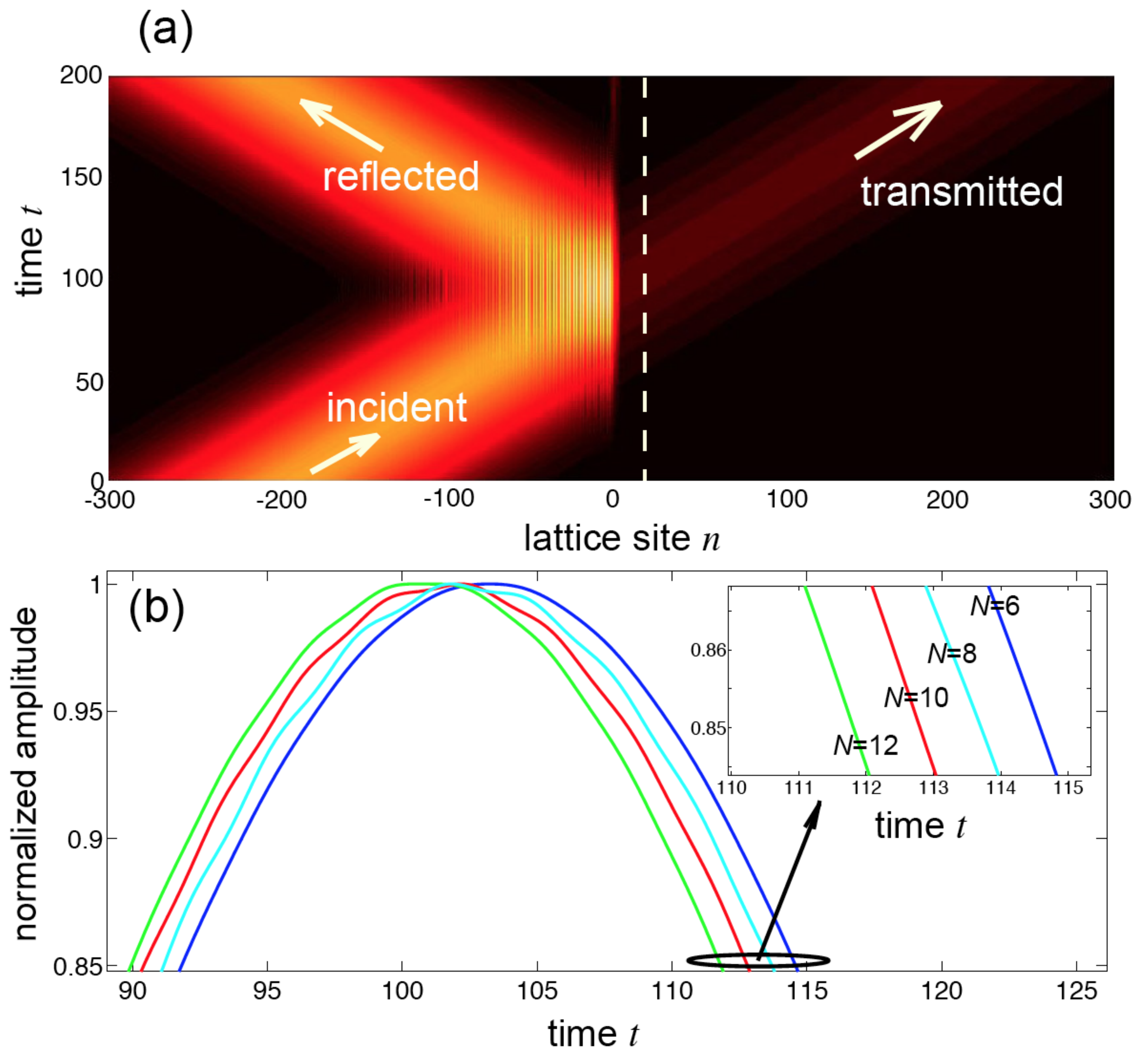}} \caption{ \small
Scattering dynamics at a NH barrier with nearest and next-to-nearest neighbor hopping amplitudes ($\kappa=1$, $t_{-2}=0.275$, $t_{-1}=-0.8667$, $t_{0}=-2.3627$, $t_1=-0.9184$ and $t_2=0.25$). (a) Space-time scattering dynamics of an incident Gaussian wave packet for barrier width $N=6$. The input wave function is $\psi_n(0)=\exp[-(n+190)^2/w_0^2-iqn] $ with carrier Bloch wave number $q=\pi /2$ and size $w_0=80$. (b) Temporal behavior of the wave function amplitude $| \psi_n(t)|$ in the site $n=20$, after the barrier [indicated by the dashed vertical curve in (a)], normalized to its maximum value, for a few increasing values of the barrier size $N$ ($N=6,8,10,12$).  {\color{black}{The barrier is located between $n=1$ and $n=N$
}}. As shown in the inset, for increasing $N$ a waveform temporal advancement is observed, with a rate $\Delta \tau / \Delta N$ of temporal advancement $\Delta \tau$ which asymptotically reaches a value close to 0.5, indicating that the phase time becomes almost independent of barrier width.}
\end{figure} 

 In the Hatano-Nelson model with only nearest-neighbor non-reciprocal hopping the tunneling dynamics, and thus the Hartman effect, can be basically reduced  to the one of an Hermitian barrier with on-site potential $t_0$ and effective nearest-neighbor reciprocal coupling $ \sqrt{t_1t_{-1}}$. This can be readily proven after a non-unitary gauge transformation of the wave function $\psi_n$, which makes the hopping reciprocal in the barrier region  (see for instance \cite{N5,K4}). This is possible because the generalized Brillouin zone in the Hatano-Nelson model is a circle, i.e. the imaginary part of the Bloch wave number $k$ is homogeneous. However, for long-range non-reciprocal hopping the generalized Brillouin zone in not anymore a circle \cite{N5,N8,N11}, and the tunneling dynamics cannot be anymore reduced to the one of an Hermitian barrier. For example, the NH Hartman effect can be observed by generalizing the Hatano-Nelson model including next-to-nearest hopping amplitudes, corresponding to $M=1$ and $s=r=2$. As an illustrative example, let us consider the NH barrier corresponding to the hopping amplitudes $t_{-2}=0.275$, $t_{-1}=-0.8667$, $t_{0}=-2.3627$, $t_1=-0.9184$ and $t_2=0.25$. {\color{black}{For such parameter values, the energy spectrum of the Hamiltonian in the barrier region under open boundaries, i.e. when isolated from the two leads, is entirely real}}. For an incident wave with an energy $E_0$ close to zero ($q \simeq \pi/2, E_0 \simeq 0$), the four roots $\beta_l$ of the determinantal equation $H(\beta)-E_0=0$ are real, with $\beta_s=\beta_2 \sim 0.85$. Figure 4 shows the spectral transmittance $T=|t|^2$ and tunneling phase time $\tau$ versus barrier size $N$, as obtained by numerically solving the matching conditions at the junctions [Eqs.(A.11-A.19) in Appendix A], for an incidence wave with zero energy ($q= \pi/2, E_0=0$), clearly indicating the existence of the NH Hartman effect. The scattering dynamics of a spatially-broad Gaussian wave packet across the NH barres, for a few increasing values of the $N$, is shown in Fig.5.The rate $ \Delta \tau / \Delta N$ of temporal advancement of the wave amplitudes as $N$ in increased, shown in Fig.5(b), becomes close to 0.5 as $N$ is increased, which is a clear signature of the NH Hartman effect.\\
 
  \begin{figure}[h]
\centerline{\includegraphics[width=8.2cm]{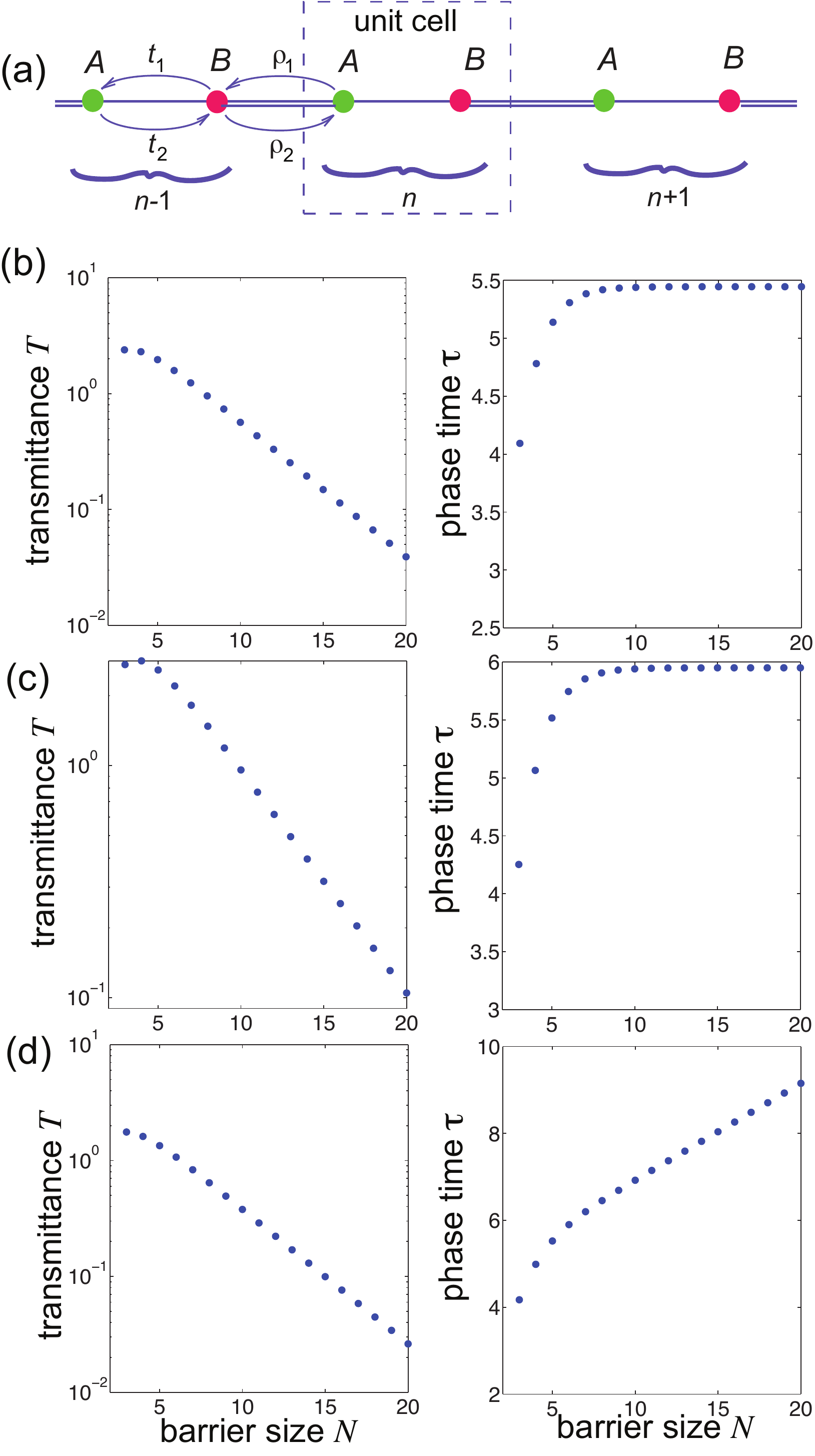}} \caption{ \small
(a) Schematic of the NH Rice-Mele superlattice. The onsite potentials in sublattices A and B are $\Delta_A$ and $\Delta_B$, respectively, while the left/right intra-dimer and inter-dimer hopping amplitudes are $t_{1,2}$ and $\rho_{1,2}$, respectively. (b-d) Behavior of the transmittance $T=|\mathcal{T}|^2$ on a log scale (left panels) and tunneling phase time $\tau$ (right panels) versus the barrier size $N$ in the NH Rice-Mele barrier for 
parameter values $\kappa=1$, $t_1=0.8$, $t_2=1$, $\rho_1=0.5$, $\rho_2=0.7$ and for a few different values of the on-site potentials: (b) $\Delta_A=\Delta_B=0$; (c) $\Delta_A=0.1 i$, $\Delta_B=-0.1 i$; (d) $\Delta_A=0$, $\Delta_B=-0.1 i$. }
\end{figure} 

  \begin{figure}[h]
 \centerline{\includegraphics[width=8.6cm]{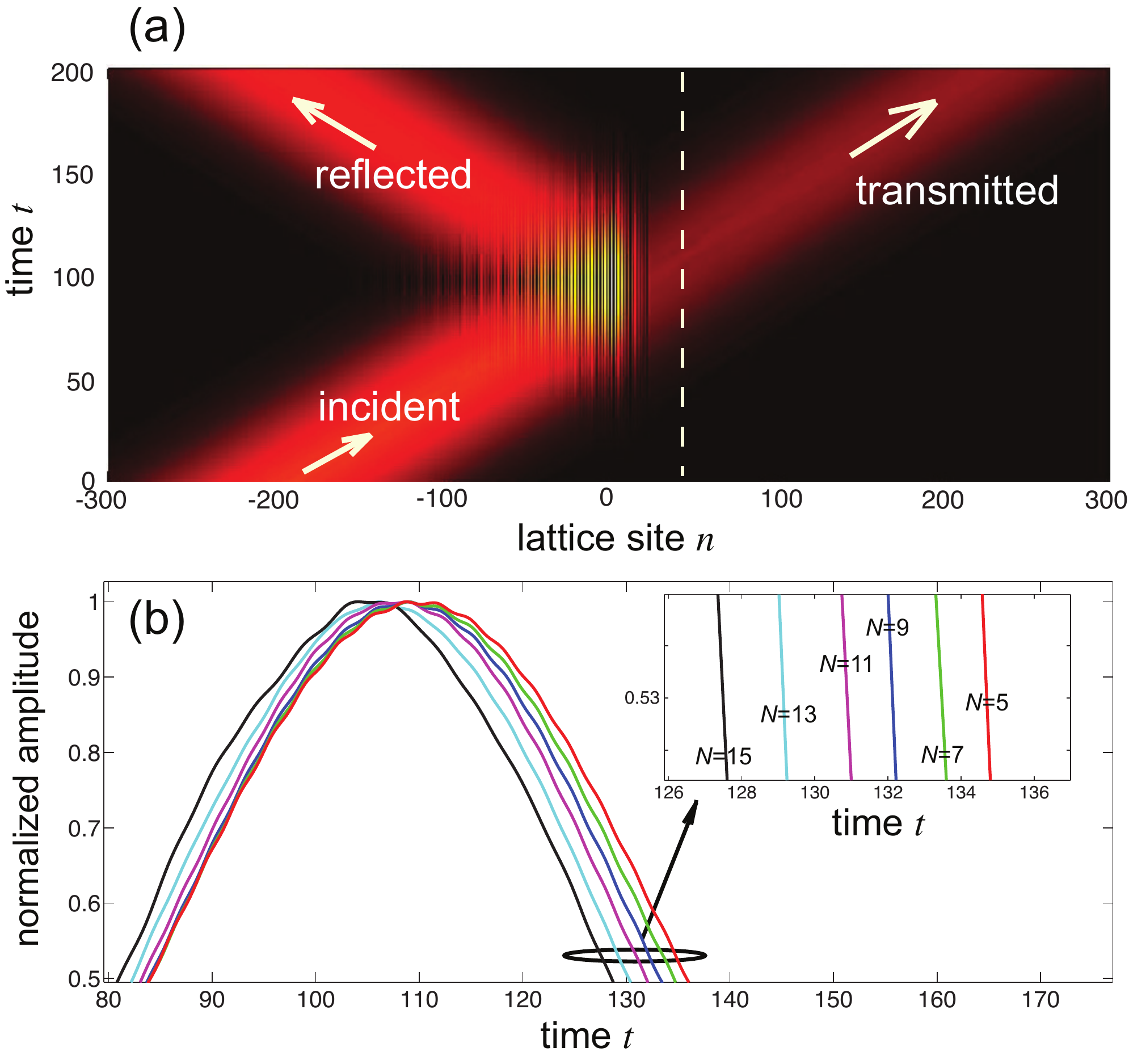}} \caption{ \small
 Scatterng dynamics at a NH Rice-Mele barrier for parameter values $\kappa=1$, $ t_1=0.8$, $t_2=1$, $\rho_1=0.5$, $\rho_2=0.7$,  and $\Delta_A=\Delta_B=0$. (a) Space-time scattering dynamics of an incident Gaussian wave packet for a barrier width $N=13$. The input wave function is $\psi_n(0)=\exp[-(n+190)^2/w_0^2-iqn] $ with carrier Bloch wave number $q=\pi /2$ and size $w_0=80$. (b) Temporal behavior of the wave function amplitude $| \psi_n(t)|$ in the site $n=40$, after the barrier [indicated by the dashed vertical curve in (a)], normalized to its maximum value, for a few increasing values of the barrier size $N$ ($N=5,7,9,11,13,15$). {\color{black}{The barrier is located between $n=1$ and $n=N$.}}  As shown in the inset, an increase $\Delta N=2$ in the barrier width $N$ (unit cell) corresponds to an advancement in time $\Delta \tau$ of $\sim 2$ for large $N$, indicating that the tunneling phase time is nearly independent of the barrier size $N$.}
\end{figure} 
{\color{black} In the above examples, to observe the Hartman effect the on-site potential energy $t_0$ has been assumed to be real. For real values of the hopping amplitudes, this ensures that the root $\beta_s(q)$ can be real over a wide range of the Bloch wave number $q$, leading to the Hartman effect. A natural and interesting question is whether the Hartman effect could be observed for a on-site potential energy $t_0$ with a non-vanishing imaginary part, for example for ${\rm Im}(t_0)<0$ corresponding to a dissipative on-site potential energy. In this case, stationarity of the phase of $\beta_s(q)$ cannot be anymore realized over a wide range of the Bloch wave number $q$, however it can occur at isolated values of $q$. This means that, for a NH barrier with dissipative on-site energy, the Hartman effect can be rather generally observed only for isolated energies of the incoming wave. For example, for a NH barrier with $r=s=2$ and for parameter values $t_{-1}=t_1=\kappa=1$, $t_{-2} =0.4$, $t_2=0.3$ and $t_0=1-5-0.3i$ the phase $\varphi_{\beta_s}(q)$ of $\beta_s(q)$ shows an isolated stationary point at $q=q_0 \simeq 1.7$, so that the Hartman effect should be observable only for an incoming wave with energy $E=2 \kappa \cos q_0 \simeq -0.13$.} 

 \subsection{The non-Hermitian Rice-Mele model}
 As a second illustrative model, we consider a NH extension of the Rice-Mele model \cite{RM1,RM2,RM3,RM4,RM5}, which is a dimeric lattice corresponding to $M=2$ sites ($A_1=$A and $A_2=$B) in each unit cell. A schematic of the Rice-Mele model is shown in Fig.6(a). The $2 \times 2$ Bloch Hamiltonian $H(\beta)$ for this model reads
 \begin{equation}
 H(\beta)= \left(
 \begin{array}{cc}
 \Delta_A & t_1+\rho_2/ \beta \\
 t_2+ \rho_1 \beta & \Delta_B
 \end{array}
 \right)
 \end{equation}
 where $\Delta_{A}$, $\Delta_B$ are the (generally complex) on-site potential energies in the two sublattices A and B, respectively, $t_{1,2}$ are the left/right intradimer hopping amplitudes and $\rho_{1,2}$ are the left/right interdimer hopping amplitudes.  We note that, for $\Delta_A=\Delta_B=0$, this model reduces to the NH extension of the Su-Schrieffer-Heeger (SSH) model with chiral symmetry, studied in recent works in the context of the NH skin effect and bulk-edge correspondence (see e.g. \cite{N5}), while for reciprocal couplings $t_1=t_2$, $\rho_1=\rho_2$ and complex on-site potential energies $\Delta_{A,B}= \pm i \delta$, this model reduces to the PT-symmetric SSH model \cite{Schomerus}. {\color{black}{In both cases we consider parameter values such that, under open boundaries, the energy spectrum of the Hamiltonian is entirely real.}}
 The values $\beta_1$ and $\beta_2$ entering in the scattering solution Eq.(4) are obtained from the determinantal equation ${\rm det} (H(\beta)-E_0)=0$, i.e. they are the roots of the second-order algebraic equation
 \begin{equation}
 t_1 \rho_1 \beta^2+[t_1 t_2+\rho_1 \rho_2-(E_0-\Delta_A)(E_0-\Delta_B)] \beta+\rho_2 t_2=0
 \end{equation}
 while the eigenvectors $U^{(1,2)}$ read
 \begin{equation}
 U^{(1)}= \left(
 \begin{array}{c}
 t_1+ \rho_2 / \beta_1 \\
 E-\Delta_A
 \end{array}
 \right) \;, \;\; 
  U^{(2)}= \left(
 \begin{array}{c}
 t_1+ \rho_2 / \beta_2 \\
  E-\Delta_A
 \end{array}
 \right).
 \end{equation}
 The spectral reflection and transmission amplitudes $\mathcal{R}(q)$ and $\mathcal{T}(q)$ are obtained by imposing the matching conditions of the scattering solution at the junctions, i.e. by imposing the following equations
 \begin{eqnarray}
 E_0 \psi_0 & = & \kappa \psi_{-1}+\kappa \psi_1^{(A)} \\
 E_0 \psi_1^{(A)} & = & \Delta_A \psi_1^{(A)} + \kappa \psi_0+t_1 \psi_1^{(B)}\\
  E_0 \psi_N^{(B)} & = & \Delta_B \psi_N^{(B)} + \kappa \psi_{N+1}+t_2 \psi_N^{(A)}\\
  E_0 \psi_{N+1} & = & \kappa \psi_{N+2}+ \kappa \psi_{N}^{(B)}.
 \end{eqnarray}
This yields the following linear inhomogeneous system in the variables $\mathcal{R}(q)$, $G_1$, $G_2$ and $\mathcal{T}(q)$
 \begin{equation}
 \mathcal{M} \left(
 \begin{array}{c}
 \mathcal{R}(q) \\
 G_1 \\
 G_2 \\
 \mathcal{T}(q)
 \end{array}
  \right)
  = \left(
 \begin{array}{c}
 E_0- \kappa \exp(iq)  \\
 \kappa \\
 0 \\
 0
 \end{array}
  \right)
 \end{equation}
 with matrix $\mathcal{M}$ given by
 
 \begin{strip}
 \begin{equation}
 \mathcal{M}= \left(
 \begin{array}{cccc}
- \kappa \exp(iq) & \kappa (t_1 \beta_1+ \rho_2) & \kappa (t_1 \beta_2+ \rho_2) & 0\\
- \kappa & \rho_2(E_0-\Delta_A) & \rho_2(E_0-\Delta_A)  & 0 \\
0 & \rho_1 \beta_1^N (t_1 \beta_1+ \rho_2) & \rho_1 \beta_2^N (t_1 \beta_2+ \rho_2) & - \kappa \\
0 & \kappa (E_0-\Delta_A) \beta_1^N & \kappa (E_0-\Delta_A) \beta_2^N & - \kappa \exp(iq)
 \end{array}
 \right).
 \end{equation}
 \end{strip}

 According to the general result shown in Sec.3, the NH Hartman effect is observed provided that the phase of the root $\beta_s=\beta_1$ to Eq.(17)  does not depend on $q$. This condition can be satisfied when the intra- and inter-dimer hopping amplitudes $t_{1,2}$, $\rho_{1,2}$ are real and the on-site potential energies  are either real or satisfy the condition $\Delta_A=\Delta_B^*$, which basically corresponds PT symmetry in the system. Hence, like in continuous models of the NH Hartman effect \cite{T42,T43}, PT symmetry ensures the independence of the tunneling phase time on barrier width for long barriers. However, the NH Hartman effect can be observed even without any complex on-site potential, for example by assuming $\Delta_A=\Delta_B=0$, as a result of non-reciprocal hopping in the lattice. Figure 6(b-d) shows an example of numerically-computed spectral transmittance and tunneling phase time in a NH Rice-Mele barrier versus barrier size $N$ for an incoming wave with Bloch wave number $q_0= \pi/2$, illustrating different tunneling regimes. In particular, note that the Hartman effect is observed in Figs.6(b) and (c), where the condition $\Delta_A=\Delta_B^*$ is satisfied, but not in Fig.6(d). The scattering dynamics of a spatially-broad Gaussian wave packet across the NH barrier,  showing the Hartman effect and corresponding to parameter values of Fig.6(b), is shown in Fig.7. Figure 7(a) depicts a typical space-time map of wave packet scattering for a barrier comprising $N=13$ unit cells,  while Fig.7(b) shows the temporal behavior of normalized waveforms observed beyond the barrier at a given lattice site ($n=40$) and for a few increasing values of $N$. Note that the rate $\Delta \tau / \Delta N$ of temporal advancement $\Delta \tau$ of the wave amplitudes, as $N$ is increased, reaches an asymptotic value close to $1$, as clearly visible in the inset of Fig.7(b). This is a clear indication of the saturation of the tunneling phase time for increasing values of $N$, i.e. of the Hartman effect .
 
 \section{Conclusion.} In this work we investigated wave scattering from a non-Hermitian potential barrier in the tight-binding picture, and derived a rather general condition for the existence of the Hartman effect, i.e.  
the independence of the phase tunneling time on barrier width. Our results extend to non-Hermitian systems on a lattice the rather paradoxical effect of wave tunneling in Hermitian systems, indicating that the Hartman effect in non-conservative systems can be observed without any special symmetry requirement, such as PT symmetry. The appearance of the NH Hartman effect has been illustrated by considering tunneling across NH barriers described by the generalized Hatano-Nelson model and by a NH extension of the Rice-Mele model.

  \appendix
  \renewcommand{\theequation}{A.\arabic{equation}}
\setcounter{equation}{0}
\begin{strip}
  \section{Matching conditions and scattering analysis}
In this Appendix we provide the detailed form of the scattering wave equations, obtained from the matching conditions at the junctions,  in the special case where we have a single site in each unit cell of the barrier, i.e. for $M=1$. The more general case $M \geq 2$ can be handled in a similar way, as an explicit example is given in Sec. 4.2 of the main text. For $M=1$, the Hamiltonian $H(\beta)$ of the barrier superlattice is a scalar and given by the Laurent polynomial
\begin{equation}
H(\beta)=\frac{t_{-s}}{ \beta^{s}}+\frac{t_{-s+1}}{ \beta^{s-1}}+...+ t_0 + t_1 \beta+ ...+ t_r \beta^r
\end{equation}
where $t_{\mp l}$ are the left or right hopping amplitudes among sites
distant $|l|$ in the lattice and $s, r \geq 1$ are the largest orders of
left ($s$) or right ($r$) hopping. We recall that, for non-reciprocal couplings, owing to the NH skin effect the energy spectrum of the NH superlattice with Hamiltonian $H(\beta)$ strongly depends on the boundary conditions \cite{ N5,N8,N11,N15, N17,healing}. In particular, the PBC energy spectrum $E_{PBC}$ is a closed loop $\mathcal{C}_1$ in complex energy plane, which is defined by the relation $E_{PBC}=H(\beta=\exp(ik))$ with $k$ real. In other words, the $E_{PBC}$ energy spectrum is the image of the unit circle $|\beta|=1$ in complex $\beta$ plane (Brillouin zone). On the other hand, the energy spectrum $E_{OBC}$ under OBC is defined by the set of energies $E$ such that $|\beta_s|=|\beta_{s+1}|$, where $\beta_l$ ($l=1,2,3,...,r+s$) are the $(r+s)$ roots of the polynomial (determinantal equation)  $ \beta^s H(\beta)-E_0 \beta^s=0$, ordered according to Eq.(3) \cite{N11,N15,N17,healing}. In a NH system displaying the NH skin effect, the energy spectrum $E_{OBC}$ differs from $E_{PBC}$ and describes one (or a set of) open arc internal to the PBC energy spectrum loop $\mathcal{C}_1$. The loci of $|\beta_s|=|\beta_{s+1}|$  in complex $\beta$ plane defines the generalized Brillouin zone, which in not the unit circle as a signature of the NH skin effect.\\
 For the scattering problem of the superlattice barrier attached to the two Hermitian tight-binding leads, for an incoming wave with wave number $q$ and energy $E_0=2 \kappa \cos q$, incident on the left side of the barrier, the scattering eigenfunction of the system at energy $E_0$ reads can be searched in the form

\begin{equation}
\psi_n= \left\{
\begin{array}{lr}
\exp(-iqn)+\mathcal{R}(q) \exp(iqn) & n \leq 0 \\
\sum_{l=1}^{s+r}G_l  \beta_l^n & 1 \leq n \leq N \\
\mathcal{T}(q) \exp[-iq(n-N-1)] & n \geq N+1\\
\end{array}
\right.
\end{equation}
where $\beta_1$, $\beta_2$,..., $\beta_{s+r}$ are the $(r+s)$ roots of the polynomial (determinantal equation)  $ \beta^s H(\beta)-E_0 \beta^s=0$, ordered according to Eq.(3).
The matching conditions at the junctions are obtained by imposing that the Ansatz (A.2) satisfies the stationary Schr\"odinger equation at the lattice sites $n=0,1,2,.., s$ and $n=N-r+1,N-r+2,...,N,N+1$, i.e. 
\begin{eqnarray}
E_0 \psi_0 & = & \kappa( \psi_{-1}+ \psi_1) \\
(E_0-t_0) \psi_1 & = & \kappa \psi_0+t_1 \psi_2+t_2 \psi_3+...+ t_r \psi_{r+1} \\
(E_0-t_0) \psi_2 & = & t_{-1} \psi_1+t_1 \psi_3+t_2 \psi_4+...+ t_r \psi_{r+2} \\
... & ...& ... \nonumber \\
(E_0-t_0) \psi_{s} & = & t_{-s+1} \psi_1+t_{-s+2} \psi_2+...+ t_r \psi_{r+s} \\
(E_0-t_0) \psi_{N-r+1} & = & t_{-s} \psi_{N-r-s+1}+t_{-s+1} \psi_{N-r-s+2}+ t_{r-1} \psi_{N} \\
... & ...& ... \nonumber \\
(E_0-t_0) \psi_{N-1} & = & t_{-s} \psi_{N-s-1}+t_{-s+1} \psi_{N-s}+...+ t_{-1} \psi_{N-2} +  t_1 \psi_N   \\
(E_0-t_0) \psi_{N} & = & t_{-s} \psi_{N-s}+t_{-s+1} \psi_{N-s+1}+...+ t_{-1} \psi_{N-1} +  \kappa \psi_{N+1} \\
E_0 \psi_{N+1} & = &\kappa  \psi_{N+2} + \kappa \psi_{N}.
\end{eqnarray}
where we assumed $N>(r+s)$.
This yields the following set of $(s+r+2)$ linear inhomogeneous equations 
\begin{eqnarray}
E_0 [1+\mathcal{R}(q)] & =  & \kappa [ \exp(iq)+\mathcal{R}(q) \exp(-iq)] +\kappa  \sum_{l=1}^{r+s} G_l  \beta_l \\
(E_0-t_0) \sum_{l=1}^{r+s} G_l  \beta_l & = &  \kappa [1+\mathcal{R}(q)]+t_1  \sum_{l=1}^{r+s} G_l  \beta_l^2 + t_2  \sum_{l=1}^{r+s} G_l  \beta_l^3+...+ t_r  \sum_{l=1}^{r+s} G_l  \beta_l^{r+1} \\
(E_0-t_0) \sum_{l=1}^{r+s} G_l  \beta_l^2 & = &  t_{-1}  \sum_{l=1}^{r+s} G_l  \beta_l+t_1  \sum_{l=1}^{r+s} G_l  \beta_l^3+ t_2  \sum_{l=1}^{r+s} G_l  \beta_l^4+...+ t_r  \sum_{l=1}^{r+s} G_l  \beta_l^{r+2} \\
... & ... & ... \nonumber \\
(E_0-t_0) \sum_{l=1}^{r+s} G_l  \beta_l^s & = &  t_{-s+1}  \sum_{l=1}^{r+s} G_l  \beta_l+t_{-s+2}  \sum_{l=1}^{r+s} G_l  \beta_l^2+...+ t_r  \sum_{l=1}^{r+s} G_l  \beta_l^{s+r} \\
(E_0-t_0) \sum_{l=1}^{r+s} G_l  \beta_l^{N-r+1} & = &  t_{-s}  \sum_{l=1}^{r+s} G_l  \beta_l^{N-s-r+1}+t_{-s+1}  \sum_{l=1}^{r+s} G_l  \beta_l^{N-r-s+2}+ ...+ t_{r -1} \sum_{l=1}^{r+s} G_l  \beta_l^{N} \\
... & ... & ... \\
(E_0-t_0) \sum_{l=1}^{r+s} G_l  \beta_l^{N-1} & = &  t_{-s}  \sum_{l=1}^{r+s} G_l  \beta_l^{N-s-1}+t_{-s+1}  \sum_{l=1}^{r+s} G_l  \beta_l^{N-s}+ ...+ t_{1} \sum_{l=1}^{r+s} G_l  \beta_l^{N} \\
(E_0-t_0) \sum_{l=1}^{r+s} G_l  \beta_l^{N} & = &  t_{-s}  \sum_{l=1}^{r+s} G_l  \beta_l^{N-s}+t_{-s+1}  \sum_{l=1}^{r+s} G_l  \beta_l^{N-s+1}+ ...+t_{-1}  \sum_{l=1}^{r+s} G_l  \beta_l^{N-1}+ \kappa \mathcal{T}(q) \\
 E_0 \mathcal{T}(q) & = & \kappa \mathcal{T}(q) \exp(-iq)+ \kappa \sum_{l=1}^{r+s} G_l \beta_l^N,
\end{eqnarray}
which can be solved to determine the amplitudes $G_1,G_2,...,G_{s+r}$ of evanescent waves in the barrier region, and the spectral transmission and refection coefficients $\mathcal{T}(q)$ and $\mathcal{R}(q)$ for a given value of the Bloch wave number $q$, i.e. energy $E_0=2 \kappa \cos q$ of the incoming wave.

    \renewcommand{\theequation}{B.\arabic{equation}}
\setcounter{equation}{0}
  \section{Asymptotic form of the spectral transmission amplitude}
  In this Appendix we prove that the spectral amplitude $\mathcal{T}(q)$, in the limit of a long barrier (i.e. in the large $N$ limit), takes the form
  \begin{equation}
  \mathcal{T}(q) =\tilde{\mathcal{T}} (q) \beta_s^{N}
  \end{equation}
  where $\tilde{\mathcal{T}}(q)$ is independent of $N$. This result holds provided that the condition $| \beta_{s}|<|\beta_{s+1}|$ is strictly satisfied, i.e. that the energy $E_{0}$ of the incoming wave does not belong to the energy spectrum $E_{OBC}$ 
  of the superlattice under OBC. \\
  To prove this statement, we should solve the inhomogeneous linear system of $(s+r+2)$ equations, obtained from the matching conditions at the junctions (see e.g. Eqs.(A.11-A.19) in Appendix A), for the unknown variables $G_1$, $G_2$,...,$G_{l+s}$, $\mathcal{R}(q)$, $\mathcal{T}(q)$ in the large $N$ limit. Since the coefficients of the linear system contain powers of $ \sim \beta_l^N$, to provide an asymptotic solution to the system in the large $N$ limit it is worth rewriting the equations after suitable scaling of the variables $G_l$ and $\mathcal{T}(q)$, so as the order of magnitude of coefficients in the resulting equations can be readily visualized in the large $N$ limit. Such a scaling analysis is similar to the one used in NH lattices displaying the NH skin effect to compute the generalized Brillouin zone (see for instance \cite{N11,N17,healing}). Specifically, let us introduce the following substitutions
  \begin{equation}
  G_l=g_l \; \; {\rm for} \;\; l=1,2,...,s, \;\; \;\; G_l=g_l \left(  \frac{\beta_s}{\beta_l}\right)^N \; \; {\rm for} \;\; l=s+1,s+2,...,s+r, \;\;\;\;\; \mathcal{T}(q) = \tilde{\mathcal{T}}(q) \beta_s^N
  \end{equation}
  and let us assume that the inequality $|\beta_s|<|\beta_{s+1}|$ is strictly satisfied. In this way, it can be readily shown that in the first $(s+1)$ equations in the system [e.g. Eqs.(A.11-A.14) in Appendix A] the coefficients of the amplitudes $g_{s+1}, g_{s+2}, .., g_{s+r}$ vanish in the large $N$ limit, since they are expressed in terms of powers  $ \sim (\beta_s / \beta_l)^N$ and $|\beta_l|> \beta_s|$ ($l=s+1,s+2,...,s+r$). Therefore, in the large $N$ limit the first set of $(s+1)$ equations of the system are decoupled from the other ones and can be solved, providing the solution to $\mathcal{R}(q)$, $g_1$, $g_2$,...,$g_s$ which are independent of $N$. The other $(r+1)$ equations of the system [e.g. Eqs.(A.15-A.19)] of Appendix A] can be then solved to determine the amplitudes $g_{s+1},g_{s+2},..., g_{s+r}$ and $\tilde{\mathcal{T}}(q)$. To this aim, it is worth dividing both sides of such set of equations by $\beta^s$. In this way, in the large $N$ limit the coefficients of the amplitudes $g_l$, with $l<s$, are vanishing while all other coefficients of $g_l$ and $\tilde{\mathcal{T}}(q)$ are independent of $N$. Therefore, the resulting solution to $\tilde{\mathcal{T}}(q)$ is independent of $N$. This proves the main result stated by Eq.(B1).
  
  \end{strip}
  
 \noindent
{\bf Disclosures.} The author declares no conflicts of interest.\\
\\
{\bf Acknowledgment.} The author acknowledges the Spanish
State Research Agency, through the Severo-Ochoa and Maria
de Maeztu Program for Centers and Units of Excellence in R\&D
(Grant No. MDM-2017-0711).\\
\\
{\bf Data Availability.} No data were generated or analyzed
in the presented research.

\end{document}